\documentclass[fleqn,usenatbib]{mnras}

\usepackage{mathptmx}

\usepackage[T1]{fontenc}

\DeclareRobustCommand{\VAN}[3]{#2}
\let\VANthebibliography\thebibliography
\def\thebibliography{\DeclareRobustCommand{\VAN}[3]{##3}\VANthebibliography}

\usepackage{graphicx}	
\usepackage{amsmath}	
\usepackage{amssymb}	

\newcommand{\msun}{{\rm M}_\odot}

\newcommand\fbinbms{5.9}
\newcommand\fbinbmserr{1.1}
\newcommand\nbinbms{20/338}
\newcommand\fbinrms{4.5}
\newcommand\fbinrmserr{0.6}
\newcommand\nbinrms{44/978}
\newcommand\fbinpb{6.6}
\newcommand\fbinpberr{1.5}

\title[Stellar rotation and binarity]{Exploring the role of binarity in the origin of the bimodal rotational velocity distribution in stellar clusters}

\author[S. Kamann et al.]{
Sebastian Kamann,$^{1}$\thanks{E-mail: s.kamann@ljmu.ac.uk}
Nate Bastian,$^{1}$
Christopher Usher,$^{2}$
Ivan Cabrera-Ziri,$^{3,1}$,
Sara Saracino$^{1}$
\\
$^{1}$Astrophysics Research Institute, Liverpool John Moores University, IC2 Liverpool Science Park, 146 Brownlow Hill, Liverpool L3 5RF, UK\\
$^{2}$Department of Astronomy, Oskar Klein Centre, Stockholm University, AlbaNova University Centre, SE-106 91 Stockholm, Sweden \\
$^{3}$ Astronomisches Rechen-Institut, Zentrum f\"ur Astronomie der Universit\"at Heidelberg, M\"onchhofstra{\ss}e 12-14, D-69120 Heidelberg, Germany\\
}

\date{Accepted XXX. Received YYY; in original form ZZZ}

\pubyear{2020}

\begin{document}
\label{firstpage}
\pagerange{\pageref{firstpage}--\pageref{lastpage}}
\maketitle

\begin{abstract}
Many young and intermediate age massive stellar clusters host bimodal distributions in the rotation rates of their stellar populations, with a dominant peak of rapidly rotating stars and a secondary peak of slow rotators.  The origin of this bimodal rotational distribution is currently debated and two main theories have been put forward in the literature.  The first is that all/most stars are born as rapid rotators and that interacting binaries brake a fraction of the stars, resulting in two populations.  The second is that the rotational distribution is a reflection of the early evolution of pre-main sequence stars, in particular, whether they are able to retain or lose their protoplanetary discs during the first few Myr.  Here, we test the binary channel by exploiting multi-epoch VLT/MUSE observations of NGC~1850, a $\sim100$~Myr massive cluster in the LMC, to search for differences in the binary fractions of the slow and fast rotating populations.  If binarity is the cause of the rotational bimodality, we would expect that the slowly rotating population should have a much larger binary fraction than the rapid rotators.  However, in our data we detect similar fractions of binary stars in the slow and rapidly rotating populations ($\fbinbms\pm\fbinbmserr$\% and $\fbinrms\pm\fbinrmserr$\%, respectively).  Hence, we conclude that binarity is not a dominant mechanism in the formation of the observed bimodal rotational distributions.
\end{abstract}




\section{Introduction}

It has recently been established that resolved young massive clusters host stars with a bimodal rotational distribution \citep[e.g.,][]{dupree17, kamann18} with a dominant peak of rapid rotators and a smaller peak of slow rotators \citep[e.g.,][]{marino18,bastian18, Kamann2020}.  This rotational bimodality manifests itself as a split main sequence \citep[][]{dantona15} and an extended main sequence turnoff (eMSTO - \citealt{bastian09}) in colour magnitude diagrams of massive clusters with ages of $\lesssim300$~Myr and $<2$~Gyr, respectively.

The origin of this rotational bimodality is still debated in the literature.  \citet{dantona15, dantona17} suggest that all/most stars (at least those above $\sim1.5\,\msun$) are born rapidly rotating, and that the minor peak consists of stars that were braked during their main sequence lifetimes.  Motivated by an observed trend among early F to B stars in the Galactic field, which show lower rotational velocities when being part of binary systems with periods of a few to $\sim 500$~ days \citep{abt04}, \citet{dantona15} propose that the slow rotators observed in clusters were braked via tides in similarly tight binary stars.  Such a scenario has also been advocated by \citet{sun19} for the specific case of NGC~2287, an open cluster in the Milky Way.

Alternatively, \citet{bastian20} argue that such a rotational bimodality may be imprinted at very early times in a cluster's life, and suggest that the key parameter is the time scale on which pre-main sequence stars lose their proto-stellar discs, with shorter (longer) removal times resulting in fast rotators (slow rotators).  These authors show that if a bimodality is set at early times it will persist for the subsequent $\sim2$~Gyr (for stars that are not magnetically braked, i.e. for $M_{\star} \gtrsim 1.5\,\msun$).

The pre-main sequence origin of the bimodality can be tested through observations of the period distributions of stars in very young clusters, across a wide range of stellar masses, and ongoing surveys (e.g., of the Galactic cluster Westerlund 2 - \citealt[][]{sabbi20}) should be able to confirm or refute this hypothesis.

Alternatively, a possible link between binarity and the origin of the bimodal distribution can be tested through time-series observations of massive clusters over a wide range of ages (from forming today to $\sim2$~Gyr).  In this scenario, tight binaries are needed to brake stars, and tight binaries are usually not destroyed via  three-body encounters, even in massive clusters \citep[e.g.][]{heggie03,ivanova05,fregeau09}.  To illustrate this, we follow \citet{ivanova05} and define the hardness $\eta$ of a binary as its binding energy relative to the typical kinetic energy of a cluster star, i.e. $\eta=G\,m_1\,m_2/(\langle m\rangle\,\sigma^2\,a)$. Here, $m_1$ and $m_2$ are the masses of the binary components, orbiting each other with a semi-major axis $a$, $\langle m \rangle$ is the average stellar mass inside the cluster, $\sigma$ its velocity dispersion, and $G$ the gravitational constant.  If we consider a circular orbit with a period of 500~days \citep[the upper limit for which a decrease in $v\sin i$ is observed in the Galactic field, cf.][]{abt04}, typical masses of $m_1=2\,M_{\odot}$ and $m_2=\langle m \rangle$, and a velocity dispersion of $\sigma=5{\rm km\,s^{-1}}$, we find $\eta\approx40$.  In other words, the binding energy of binaries suggested to be responsible for the formation of the sequence of slow rotators is at least $40\times$ higher than the kinematic energy of a typical cluster star.  It is very unlikely that such binary star is destroyed in a dynamical encounter with another cluster member\footnote{Note that hard binaries can still be destroyed via stellar evolution, e.g. following common envelope phases or supernova explosions. However, as the division into fast and slow rotators is present among unevolved stars, we can safely ignore such effects.}.  Hence, the binary scenario predicts that the population of slow rotators is expected to display a much higher binary fraction than the rapidly rotating population.

In the present work we explicitly test the interacting binary scenario using the integral field spectrograph MUSE, on the VLT, to study the stellar populations in the young LMC cluster NGC~1850 ($\sim100$~Myr).  VLT/MUSE allows us to obtain the radial velocities of thousands of cluster members per pointing, and by obtaining multiple epochs we can search for radial velocity variations which would indicate binarity.  In \citet{Kamann2020}, such an analysis has been applied to the intermediate-age cluster NGC~1846, where no differences in the binary frequencies of the fast and slowly rotating populations of eMSTO stars were found.

\section{Observations and data reduction}
\label{sec:observations}

The observations used in the present work will be presented in detail in a forthcoming paper (Kamann et al., in prep.)  Here, we provide a brief description.  We observed NGC~1850 with the MUSE integral field spectrograph \citep{muse} at the ESO Very Large Telescope in wide-field mode (WFM, programs: 0102.D-0268 and 106.216T.001, PI: Bastian).  The ground-layer adaptive optics (AO) system was used in order to improve the spatial resolution of the cubes.  

Our observations targeted two fields within the cluster.  One field was centred on the cluster centre while the second outer field was situated $\sim1\arcmin$ to the south-east of the centre, with a small overlap with the central field.  The observations were taken over sixteen nights between 2019-01-13 and 2021-02-06, with the sampling between individual periods ranging from 1~hour to 25~months. This ensures that our data are sensitive to binaries over a wide range of orbital periods. Nevertheless, we also perform a thorough investigation of the selection effects inherent in our  analysis (cf. Sect.~\ref{sec:results}, below).

During each night, one observing block (OB) was executed, consisting of $2\times400$~s exposures on the central field, followed by $3\times500$~s on the outer field.  We used versions 2.6 and 2.8.3 of the official MUSE pipeline \citep{pipeline} to create two data cubes -- one for the central and one for the outer pointing -- from the data observed within each OB.

We further gathered archival \textit{Hubble} space telescope (HST) photometry of NGC~1850, taken with the WFC3 camera during programs 14069 (PI: Bastian) and 14174 (PI: Goudfrooij). The data include images taken in six filters, F275W, F336W, F343N, F438W, F656N, and F814W. All images were analysed using \textsc{DOLPHOT}, a modified version of \textsc{HSTphot} \citep{hstphot}, following the procedure described in \citet{gossage2019} and references therein. We used the RA and Dec as well as the F656N magnitudes from the final HST catalog as a reference and extracted individual stellar spectra from the final MUSE data cubes using \textsc{PampelMuse} \citep{pampelmuse}. The reason for using the F656N magnitudes when initializing the extraction process is that the photometry in this narrow band filter is less affected by saturation of the brightest sources than the broad band filters overlapping with the MUSE spectral range.

\begin{figure}
    \centering
    \includegraphics[width=.95\linewidth]{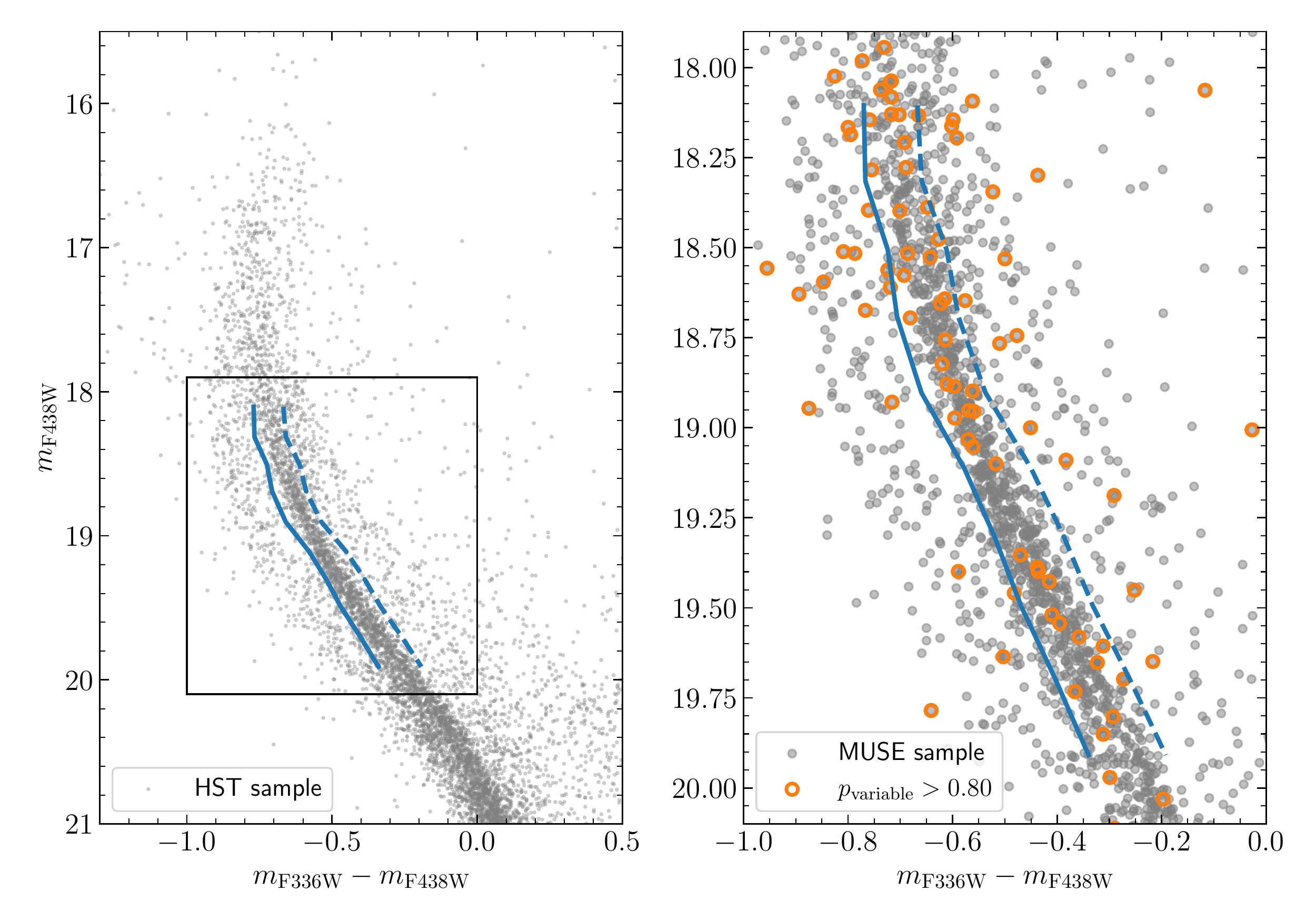}
    \caption{\textit{Left:} Colour-magnitude diagram of NGC~1850, generated from the HST photometry discussed in Sect.~\ref{sec:observations}. Blue lines indicate the colours used to separate blue and red main-sequence stars (solid) as well as red main-sequence stars and the photometric binaries (dashed). A zoom into the area highlighted by a black square is shown in the right panel. \textit{Right:} Distribution of stars with detected radial velocity variability across the split main sequences of NGC~1850. The entire MUSE sample is shown as light grey dots. Stars with high probabilities of being variable are highlighted with orange circles. Blue lines are the same as in the left panel.}
    \label{fig:binary_stars}
\end{figure}

We show the distribution of the entire HST sample in the ($m_{\rm F336W} - m_{\rm F438W}$, $m_{\rm F438W}$) colour-magnitude diagram in the left panel of Fig.~\ref{fig:binary_stars}. The split main sequence is visible between approximately $18.0 < {\rm m}_{\rm F438W} < 20$, and we show the distinction between the two branches (blue MS and red MS) as a solid (blue) line.  The separation was performed by dividing the stars in the magnitude range $18.0 < {\rm m}_{\rm F438W} < 20.0$ into bins of $0.2\,{\rm mag}$ width. In each magnitude bin, we inspected the distribution of $m_{\rm F336W} - m_{\rm F438W}$ colours, searching for the minimum between the two sequences. As the minimum was always located around the 20th percentile of the distribution, we used this value to separate the two branches. We note that \citet{correnti2017} measured a fraction of blue main sequence stars of $\sim25-30\%$ in NGC~1850, i.e. found a slightly higher value than the $20\%$ adopted in this work.

The photometric binary sequence of NGC~1850 is visible to the red of the red main sequence in Fig.~\ref{fig:binary_stars}. In order to distinguish between red main-sequence stars and photometric binaries, we only considered as part of the red main sequence the stars that lie bluewards of the dashed blue line included in Fig.~\ref{fig:binary_stars}. The line was constructed by shifting the dividing line between the two main sequences (i.e. the solid blue line, cf. Fig.~\ref{fig:binary_stars}) by an amount $\delta c$ in $m_{\rm F336W} - m_{\rm F438W}$ colour\footnote{We note that photometric binaries will not only appear redder, but also brighter compared to single stars. However, as visible in Fig.~\ref{fig:binary_stars}, the adopted approach still enables us to identify most photometric binaries.}, where $\delta c$ increased linearly from $0.1$ at $m_{\rm F438W}=18.0$ to 0.15 at $m_{\rm F438W}=20.0$. This was done in order to account for the increased photometric scatter when going towards fainter magnitudes.

\section{Stellar parameters and binarity}
\label{sec:parameters}

We used \textsc{Spexxy} \citep{spexxy} in order to analyse the extracted spectra. The code determines radial velocities as well as stellar parameters (effective temperature, metallicity) using full-spectrum fits against a library of template spectra. We used the synthetic templates presented by \citet{ferre}, which extend to effective temperatures of $T_{\rm eff}=30\,000\,{\rm K}$, higher than what is expected along the main sequence of a 100~Myr old cluster. The initial guesses required by \textsc{Spexxy} were taken from the comparison of the HST photometry to a MIST \citep{mist0,mist1,mist-rotation} isochrone, assuming an age of 100~Myr and a metallicity of $[{\rm Fe/H}]=-0.24$.  Note that in contrast to the analysis presented in \citet{Kamann2020} for NGC~1846, no line broadening was included in the fits. Given that the synthetic templates were convolved with a model for the MUSE line spread function (LSF) prior to the fits, this implies that all spectra were analysed under the assumption that $v\sin i=0$. An analysis of the stellar rotation measurements obtained from the MUSE data will be presented in a separate paper (Kamann et al., in prep.).

We used the radial velocities determined from the single-epoch spectra in order to search for potential binary stars. Note that we only search for signs of variability in this study, in order to perform a comparison between the fractions of potential binary stars on the blue and red main sequences. We make no effort to constrain the orbits of individual binary systems or the distribution of binary periods inside the cluster. Such an analysis requires additional epochs, which are only now becoming available as part of a follow-up program we are currently carrying out with MUSE. An in-depth analysis of the binary systems in NGC~1850 will be presented in a dedicated publication (Saracino et~al., subm.).

In order to detect stars showing radial velocity variability, we need to be able to distinguish between intrinsic variability and variations caused by the finite S/N of our data. To this aim, a proper knowledge of the measurement uncertainties is key. In \citet{Kamann2020}, we presented a method to calibrate the velocity uncertainties returned by the \textsc{Spexxy} code. In brief, it works by measuring the period-to-period scatter in the measured radial velocities for all stars. Then, for each star a comparison sample is created, consisting of 100 stars with similar stellar parameters according to their position in the CMD. For each star under consideration, a correction factor for the uncertainty of each measured radial velocity is calculated as the standard deviation of the normalized velocity differences of its comparison sample. For more details about the procedure, we refer to \citet{Kamann2020}. We find median uncertainties of $21.4\,{\rm km\,s^{-1}}$ and $18.3\,{\rm km\,s^{-1}}$ for stars on the red and blue main sequences, respectively. The values are considerably larger than what we found for eMSTO stars in NGC~1846 in \citet{Kamann2020}, even though the typical S/N of the spectra is higher in the present study. The reasons behind this are twofold. First, the stars under consideration in NGC~1850 are hot stars, which in the wavelength range covered by MUSE show mainly broad Balmer and Paschen lines. Second, there is strong diffuse gas emission across the observed fields of NGC~1850, which can potentially leave residuals in the extracted spectra. For this reason, we decided to mask out the wavelength ranges where gas emission is present during the analysis with \textsc{Spexxy}, including the cores of the strong hydrogen lines.

We used the method developed by \citet{Giesers2019} to convert the combined set of radial velocity measurements into a probability, $p$, for each star, that it shows radial velocity variations. The method uses the individual velocities and their uncertainties in order to calculate a reduced $\chi^2$ value for each star under the assumption it does not show radial velocity variations. The distribution of reduced $\chi^2$ values obtained this way is compared to the complementary cumulative distribution function expected for the reduced $\chi^2$ values if there were no variable stars in the sample. Essentially, for any given reduced $\chi^2$ value, the method determines the ratio between the number of stars \emph{detected} above this value and the number of stars \emph{expected} above this value in the absence of variable stars. Based on this comparison, each star is assigned a probability $p$ of being variable. High ratios correspond to high $p$ values, and vice versa. For further details, we refer to \citet{Giesers2019}.

We considered each star with $p>0.8$ as a likely binary star. The distribution of those stars along the split main sequences of NGC~1850 is shown in the right panel of Fig.~\ref{fig:binary_stars}. As mentioned in Sect.~\ref{sec:observations}, we used the blue line included in Fig.~\ref{fig:binary_stars} to select two samples of blue and red main sequence stars in the magnitude range $18.0<m_{\rm F438W}<20.0$ where the two sequences can be well separated. 

It is obvious from the large radial velocity uncertainties that the method applied here is mainly sensitive to tight binaries, as these are the ones with the largest expected radial velocity variations, and also are expected to show such variations over the timescales sampled by our observations.  While we note that these are also the systems that are predicted to be most effective in braking rapidly rotating stars \citep[e.g.,][]{dantona15}, it is evident that we have to take into account the selection effects inherent to our approach. This will be done in Sect.~\ref{sec:results} below. However, we note that selection effects are expected to be equivalent for the two populations (slow/blue-MS and fast/red-MS), hence a direct comparison is still possible.

\section{Results}
\label{sec:results}

We find comparable discovery fractions for both sequences, with \nbinbms\ ($\fbinbms\pm\fbinbmserr\%$) of the blue main sequence stars (which consist presumably of slow rotators) and \nbinrms\ ($\fbinrms\pm\fbinrmserr\%$) of the red main sequence stars (fast rotators) showing variability.  Note that the uncertainties provided for the individual discovery fractions account only for the statistical uncertainties due to the limited sample sizes. They do \emph{not} account for the binary systems that are missed due to the limited velocity accuracy and the finite sampling of the periods. That is, the true binary fractions among the red and blue MS populations are likely considerably larger than the discovery fractions reported in this work. This is also evident from the binary fraction of $\fbinpb\pm\fbinpberr\%$ that we obtain for the photometric binary stars, i.e. the stars to the red of the dashed blue line in Fig.~\ref{fig:binary_stars}. Detecting photometric binaries, i.e. systems where both stars contribute to the observed flux, via low resolution spectroscopy is challenging. The anticyclical motion of the companion star diminishes any shift observed in the spectral lines of the primary star, to the extent that the shift completely vanishes for systems composed of two equally bright stars \citep[see discussion in][]{Giesers2019,Bodensteiner2021}. Hence, it is not surprising that our discovery rate for such systems is low.

Considering photometric binaries, it is likely that some stars on the red main sequence are in fact binary systems composed of a blue main-sequence primary and a companion bright enough to shift it onto the red main sequence. Such systems could spuriously inflate our discovery fraction for the red main sequence. However, due to our aforementioned insensitivity to such systems, we do not expect this to be the case.

While we expect incompleteness to affect our discovery fractions for both populations equally, we nevertheless decided to investigate how it impacts our results. To this aim, we created mock observations in the following way. Each star with at least one MUSE radial velocity measurement was assigned a radial velocity from a Gaussian velocity distribution with a dispersion of $\sigma=5\,{\rm km\,s^{-1}}$ (as expected for NGC~1850). Then it was assumed that a fraction $f_{\rm binary}$ of the observed stars are in binary systems. Those were assigned a mass ratio $m_1/m_2$, a period $T$, and an inclination $i$. We assumed a flat distribution of mass ratios in the interval $0.4 < m_1/m_2 < 0.8$, a period distribution which was uniform in log-space and covered the range $0.2 < \log(T/{\rm d}) < 3.5$, and an isotropic distribution of inclination angles. The distributions adopted for $m_1/m_2$ and $T$ are qualitatively similar to those measured for OB stars in the Tarantula nebula \citep{sana2013,dunstall2015}. We sound a note of caution that the binary properties inside star clusters most likely depend on cluster age and stellar spectral type, hence there is no guarantee that the adopted values adequately describe the true properties of the binary population inside NGC~1850. However, observational studies of binarity in young massive star clusters of different ages are still lacking, thus we have chosen to reproduce the binary properties of the ``closest environment type'' currently available.

After we assigned each observed star a mass $m_1$, using the results from the isochrone comparison described in Sect.~\ref{sec:parameters}, and a random phase offset, we were able to calculate its projected orbital velocities at the times when we performed our measurements. Those were added to the systemic radial velocities. Finally, we created a mock data set by adding the calibrated velocity uncertainties from our true MUSE measurements to the velocities thus created. We then analysed the mock data set as described in Sect~\ref{sec:parameters}.

\begin{figure}
    \centering
    \includegraphics[width=.95\linewidth]{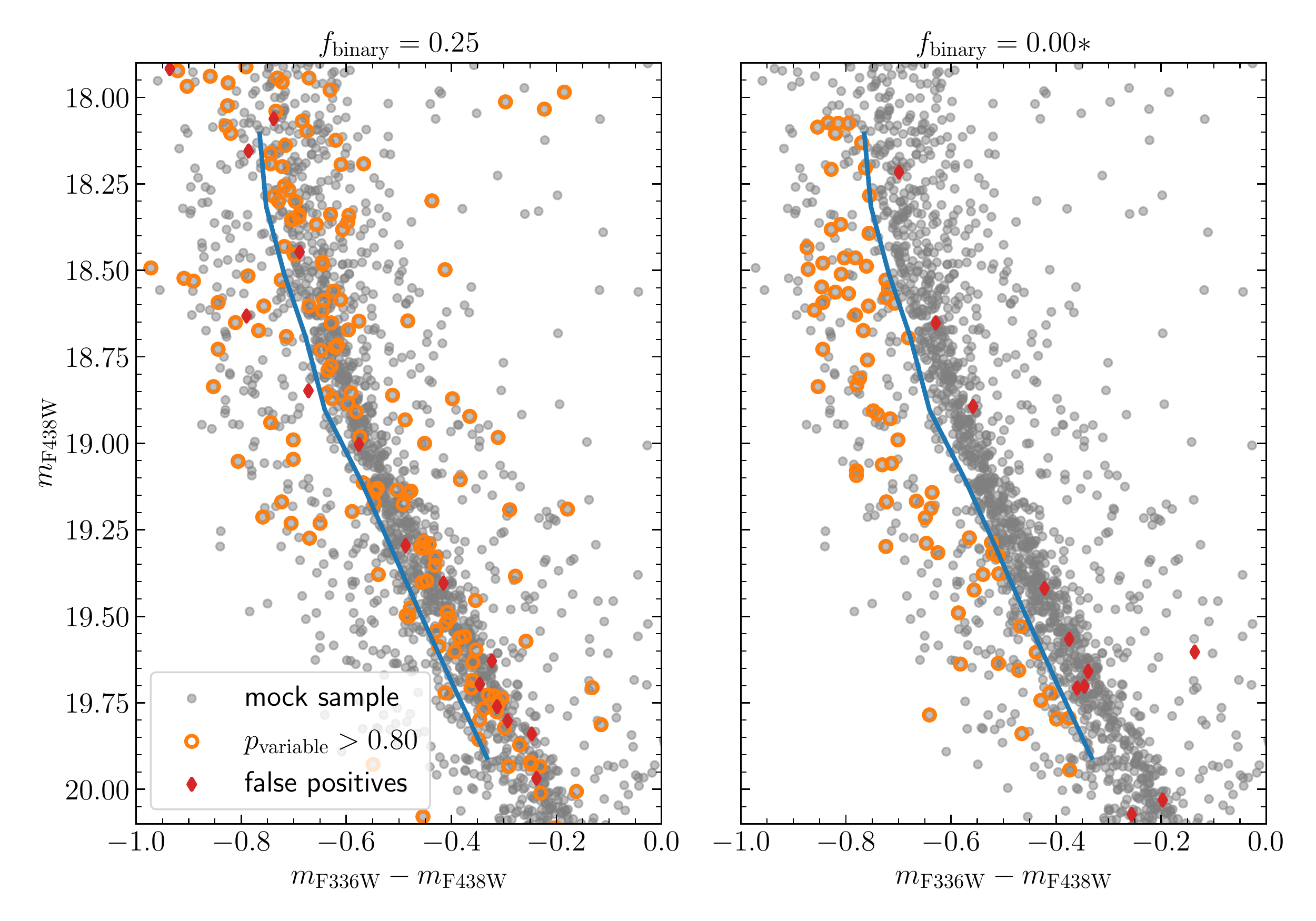}
 \caption{Results from the analysis of mock MUSE data sets for two different assumed binary fractions. Both panels show the stars present in the MUSE sample as grey dots. Stars correctly identified as binary stars in the mock analysis are highlighted as orange circles, while false positives are highlighted as red diamonds. The solid blue line indicates the assumed division between the blue and red main sequences. In the simulation shown in the left panel, we assumed a global binary fraction of $25\%$ irrespective of CMD position. On the other hand, the right panel shows the outcome of the simulation with a binary fraction that is $0\%$ everywhere but $100\%$ for blue main sequence stars.}
   \label{fig:simulation}
\end{figure}

For simplicity, we restrict ourselves to two mock simulations representative of the following scenarios. In the first one, we assume a global binary fraction of $f_{\rm binary}=0.25$\footnote{The value of $f_{\rm binary}=0.25$ was chosen as a compromise between the high binary fractions observed in young massive clusters and the low values observed in old globular clusters}, irrespective of stellar mass or CMD location. In the second simulation, we input a binary fraction of $f_{\rm binary}=0$, except for stars on the blue MS (as defined in Fig.~\ref{fig:binary_stars}), which are all assigned to a binary system (using the same period and mass ratio distributions as described above). The outcome of these two simulations is shown in Fig.~\ref{fig:simulation}. Note that we also highlight false positives in Fig.~\ref{fig:simulation}, i.e. stars that are spuriously considered to be part of binary systems due to the finite accuracy of our methodology.

The outcome of the simulation with a global binary fraction of $0.25$ is shown in the left panel of Fig.~\ref{fig:simulation}. Applying the same analysis as for the actual data yields recovery fractions of 95/1070 ($8.9 \pm 0.7\%$) and 34/367 ($9.3 \pm 1.3\%$) for the red and blue sequences. This is slightly higher than the discovery fractions we found for the actual data. On the other hand, in the extreme case shown in the right panel of Fig.~\ref{fig:simulation}, where every star on the blue sequence is considered as part of a binary system, we obtain discovery fractions of 8/1070 ($0.7 \pm 0.2\%$) and 74/367 ($20.2 \pm 1.8\%$) for the red and blue sequences, inconsistent with the results derived from the actual data. Note that our simulation is conservative in the sense that the binary systems adopted for stars on the blue MS include orbital periods $T>500\,{\rm d}$, which are unlikely to brake fast spinning stars. In summary, the outcome of these simulations support the evidence for comparable binary properties among the red and blue MS stars in NGC~1850.

We can use the simulations to estimate the completeness of our binary detection as a function of period. Averaged over all companion masses and inclinations, we find that we detect $\sim80\%$ of the binaries on the red main sequence with simulated periods of $T<5\,{\rm d}$. The efficiency drops to $\sim 60\%$ for periods of $10\,{\rm d}$ and $\sim 15\%$ for periods of $100\,{\rm d}$. For blue main sequence stars, the recovery fraction is about $5\%$ higher for the period range $10<P/{\rm d}<100$, a result of the slightly smaller radial velocity uncertainties of the blue main-sequence stars (cf. Sect.~\ref{sec:parameters}). Insofar, our simulations also confirm that there are no significant biases in our analysis that would cause us to miss more binaries among blue main sequence stars than among red main sequence stars. Instead, we are slightly biased against detecting red main sequence binaries. Arguably, this only applies if our assumption that binary systems in either main sequence share the same physical properties, in terms of mass ratio, inclination, or period distribution.

\section{Conclusions}
\label{sec:conclusions}

The similarity in the binary fractions of the red and blue main sequences is in stark contrast with expectations for the theory that the bimodal rotational distribution is due to braking by binaries.  In such a case, we would expect the blue/slow rotators to show a high binary fraction while the red/fast rotating population should show a much lower binary frequency.  We conclude that the observations of NGC~1850 are inconsistent with the idea that the observed bimodal rotational distribution of stars within clusters is due to binary interactions.

These conclusions are similar to those reached by \citet{Kamann2020} who studied the intermediate age ($\sim1.5$~Gyr) LMC cluster, NGC~1846.  These authors split their sample of eMSTO stars into slow and fast rotating sub-populations and used the same method adopted here to estimate the binary fractions of each population.  Similar to that found here, the authors report an observed binary fraction of $5.4\pm1.4$\% and $7.3\pm1.5$\% for the fast and slowly rotating subsamples of NGC~1846, respectively.

The similarity in the binary fractions amongst slow and rapidly rotating stars, in two massive clusters (with ages of $\sim100$~Myr and $\sim1.5$~Gyr) shows that interacting binaries are unlikely to play a dominant role in the origin of the bimodal rotational distribution of stars seen in clusters.  Future work on the rotational properties of pre-main sequence stars across a wide stellar mass range will be able to test the alternative theory that the bimodality is present at very young ages, and may be driven by the retention or loss of the protostellar discs around young stars.

\section*{Acknowledgements}
We thank the anonymous referee for a helpful report. SK gratefully acknowledges funding from UKRI in the form of a Future Leaders Fellowship (grant no. MR/T022868/1).
SK, NB, and SS gratefully acknowledge financial support from the European Research Council (ERC-CoG-646928, Multi-Pop). 
NB also recognises financial support from the Royal Society in the form of a University Research Fellowship.
CU acknowledges the support of the Swedish Research Council, Vetenskapsr{\aa}det.
Based on observations collected at the European Organisation for Astronomical Research in the Southern Hemisphere under ESO programmes 0102.D-0268(A) and 106.216T.001.
Based on observations made with the NASA/ESA Hubble Space Telescope and obtained from the Hubble Legacy Archive, which is a collaboration between the Space Telescope Science Institute (STScI/NASA), the Space Telescope European Coordinating Facility (ST-ECF/ESA) and the Canadian Astronomy Data Centre (CADC/NRC/CSA).

\section*{Data Availability}

The data underlying this article will be shared on reasonable request to the corresponding author.



\bibliographystyle{mnras}
\bibliography{ngc1850} 

\begin{thebibliography}{}
\makeatletter
\relax
\def\mn@urlcharsother{\let\do\@makeother \do\$\do\&\do\#\do\^\do\_\do\%\do\~}
\def\mn@doi{\begingroup\mn@urlcharsother \@ifnextchar [ {\mn@doi@}
  {\mn@doi@[]}}
\def\mn@doi@[#1]#2{\def\@tempa{#1}\ifx\@tempa\@empty \href
  {http://dx.doi.org/#2} {doi:#2}\else \href {http://dx.doi.org/#2} {#1}\fi
  \endgroup}
\def\mn@eprint#1#2{\mn@eprint@#1:#2::\@nil}
\def\mn@eprint@arXiv#1{\href {http://arxiv.org/abs/#1} {{\tt arXiv:#1}}}
\def\mn@eprint@dblp#1{\href {http://dblp.uni-trier.de/rec/bibtex/#1.xml}
  {dblp:#1}}
\def\mn@eprint@#1:#2:#3:#4\@nil{\def\@tempa {#1}\def\@tempb {#2}\def\@tempc
  {#3}\ifx \@tempc \@empty \let \@tempc \@tempb \let \@tempb \@tempa \fi \ifx
  \@tempb \@empty \def\@tempb {arXiv}\fi \@ifundefined
  {mn@eprint@\@tempb}{\@tempb:\@tempc}{\expandafter \expandafter \csname
  mn@eprint@\@tempb\endcsname \expandafter{\@tempc}}}

\bibitem[\protect\citeauthoryear{{Abt} \& {Boonyarak}}{{Abt} \&
  {Boonyarak}}{2004}]{abt04}
{Abt} H.~A.,  {Boonyarak} C.,  2004, \mn@doi [\apj] {10.1086/423795}, \href
  {https://ui.adsabs.harvard.edu/abs/2004ApJ...616..562A} {616, 562}

\bibitem[\protect\citeauthoryear{{Allende Prieto}, {Koesterke}, {Hubeny},
  {Bautista}, {Barklem}  \& {Nahar}}{{Allende Prieto} et~al.}{2018}]{ferre}
{Allende Prieto} C.,  {Koesterke} L.,  {Hubeny} I.,  {Bautista} M.~A.,
  {Barklem} P.~S.,   {Nahar} S.~N.,  2018, \mn@doi [\aap]
  {10.1051/0004-6361/201732484}, \href
  {https://ui.adsabs.harvard.edu/abs/2018A&A...618A..25A} {618, A25}

\bibitem[\protect\citeauthoryear{{Bacon} et~al.,}{{Bacon} et~al.}{2010}]{muse}
{Bacon} R.,  et~al., 2010, in Ground-based and Airborne Instrumentation for
  Astronomy III. p. 773508, \mn@doi{10.1117/12.856027}

\bibitem[\protect\citeauthoryear{{Bastian} \& {de Mink}}{{Bastian} \& {de
  Mink}}{2009}]{bastian09}
{Bastian} N.,  {de Mink} S.~E.,  2009, \mn@doi [\mnras]
  {10.1111/j.1745-3933.2009.00696.x}, \href
  {https://ui.adsabs.harvard.edu/abs/2009MNRAS.398L..11B} {398, L11}

\bibitem[\protect\citeauthoryear{{Bastian}, {Kamann}, {Cabrera-Ziri}, {Georgy},
  {Ekstr{\"o}m}, {Charbonnel}, {de Juan Ovelar}  \& {Usher}}{{Bastian}
  et~al.}{2018}]{bastian18}
{Bastian} N.,  {Kamann} S.,  {Cabrera-Ziri} I.,  {Georgy} C.,  {Ekstr{\"o}m}
  S.,  {Charbonnel} C.,  {de Juan Ovelar} M.,   {Usher} C.,  2018, \mn@doi
  [\mnras] {10.1093/mnras/sty2100}, \href
  {https://ui.adsabs.harvard.edu/abs/2018MNRAS.480.3739B} {480, 3739}

\bibitem[\protect\citeauthoryear{{Bastian}, {Kamann}, {Amard}, {Charbonnel},
  {Haemmerl{\'e}}  \& {Matt}}{{Bastian} et~al.}{2020}]{bastian20}
{Bastian} N.,  {Kamann} S.,  {Amard} L.,  {Charbonnel} C.,  {Haemmerl{\'e}} L.,
    {Matt} S.~P.,  2020, \mn@doi [\mnras] {10.1093/mnras/staa1332}, \href
  {https://ui.adsabs.harvard.edu/abs/2020MNRAS.495.1978B} {495, 1978}

\bibitem[\protect\citeauthoryear{{Bodensteiner} et~al.,}{{Bodensteiner}
  et~al.}{2021}]{Bodensteiner2021}
{Bodensteiner} J.,  et~al., 2021, arXiv e-prints, \href
  {https://ui.adsabs.harvard.edu/abs/2021arXiv210413409B} {p. arXiv:2104.13409}

\bibitem[\protect\citeauthoryear{{Choi}, {Dotter}, {Conroy}, {Cantiello},
  {Paxton}  \& {Johnson}}{{Choi} et~al.}{2016}]{mist1}
{Choi} J.,  {Dotter} A.,  {Conroy} C.,  {Cantiello} M.,  {Paxton} B.,
  {Johnson} B.~D.,  2016, \mn@doi [\apj] {10.3847/0004-637X/823/2/102}, \href
  {https://ui.adsabs.harvard.edu/abs/2016ApJ...823..102C} {823, 102}

\bibitem[\protect\citeauthoryear{{Correnti}, {Goudfrooij}, {Bellini}, {Kalirai}
   \& {Puzia}}{{Correnti} et~al.}{2017}]{correnti2017}
{Correnti} M.,  {Goudfrooij} P.,  {Bellini} A.,  {Kalirai} J.~S.,   {Puzia}
  T.~H.,  2017, \mn@doi [\mnras] {10.1093/mnras/stx010}, \href
  {https://ui.adsabs.harvard.edu/abs/2017MNRAS.467.3628C} {467, 3628}

\bibitem[\protect\citeauthoryear{{D'Antona}, {Di Criscienzo}, {Decressin},
  {Milone}, {Vesperini}  \& {Ventura}}{{D'Antona} et~al.}{2015}]{dantona15}
{D'Antona} F.,  {Di Criscienzo} M.,  {Decressin} T.,  {Milone} A.~P.,
  {Vesperini} E.,   {Ventura} P.,  2015, \mn@doi [\mnras]
  {10.1093/mnras/stv1794}, \href
  {https://ui.adsabs.harvard.edu/abs/2015MNRAS.453.2637D} {453, 2637}

\bibitem[\protect\citeauthoryear{{D'Antona}, {Milone}, {Tailo}, {Ventura},
  {Vesperini}  \& {di Criscienzo}}{{D'Antona} et~al.}{2017}]{dantona17}
{D'Antona} F.,  {Milone} A.~P.,  {Tailo} M.,  {Ventura} P.,  {Vesperini} E.,
  {di Criscienzo} M.,  2017, \mn@doi [Nature Astronomy]
  {10.1038/s41550-017-0186}, \href
  {https://ui.adsabs.harvard.edu/abs/2017NatAs...1E.186D} {1, 0186}

\bibitem[\protect\citeauthoryear{{Dolphin}}{{Dolphin}}{2000}]{hstphot}
{Dolphin} A.~E.,  2000, \mn@doi [\pasp] {10.1086/316630}, \href
  {https://ui.adsabs.harvard.edu/abs/2000PASP..112.1383D} {112, 1383}

\bibitem[\protect\citeauthoryear{{Dotter}}{{Dotter}}{2016}]{mist0}
{Dotter} A.,  2016, \mn@doi [\apjs] {10.3847/0067-0049/222/1/8}, \href
  {https://ui.adsabs.harvard.edu/abs/2016ApJS..222....8D} {222, 8}

\bibitem[\protect\citeauthoryear{{Dunstall} et~al.,}{{Dunstall}
  et~al.}{2015}]{dunstall2015}
{Dunstall} P.~R.,  et~al., 2015, \mn@doi [\aap] {10.1051/0004-6361/201526192},
  \href {https://ui.adsabs.harvard.edu/abs/2015A&A...580A..93D} {580, A93}

\bibitem[\protect\citeauthoryear{{Dupree} et~al.,}{{Dupree}
  et~al.}{2017}]{dupree17}
{Dupree} A.~K.,  et~al., 2017, \mn@doi [\apjl] {10.3847/2041-8213/aa85dd},
  \href {https://ui.adsabs.harvard.edu/abs/2017ApJ...846L...1D} {846, L1}

\bibitem[\protect\citeauthoryear{{Fregeau}, {Ivanova}  \& {Rasio}}{{Fregeau}
  et~al.}{2009}]{fregeau09}
{Fregeau} J.~M.,  {Ivanova} N.,   {Rasio} F.~A.,  2009, \mn@doi [\apj]
  {10.1088/0004-637X/707/2/1533}, \href
  {https://ui.adsabs.harvard.edu/abs/2009ApJ...707.1533F} {707, 1533}

\bibitem[\protect\citeauthoryear{{Giesers} et~al.,}{{Giesers}
  et~al.}{2019}]{Giesers2019}
{Giesers} B.,  et~al., 2019, \mn@doi [\aap] {10.1051/0004-6361/201936203},
  \href {https://ui.adsabs.harvard.edu/abs/2019A&A...632A...3G} {632, A3}

\bibitem[\protect\citeauthoryear{{Gossage} et~al.,}{{Gossage}
  et~al.}{2019a}]{gossage2019}
{Gossage} S.,  et~al., 2019a, \mn@doi [\apj] {10.3847/1538-4357/ab5717}, \href
  {https://ui.adsabs.harvard.edu/abs/2019ApJ...887..199G} {887, 199}

\bibitem[\protect\citeauthoryear{{Gossage} et~al.,}{{Gossage}
  et~al.}{2019b}]{mist-rotation}
{Gossage} S.,  et~al., 2019b, \mn@doi [\apj] {10.3847/1538-4357/ab5717}, \href
  {https://ui.adsabs.harvard.edu/abs/2019ApJ...887..199G} {887, 199}

\bibitem[\protect\citeauthoryear{{Heggie} \& {Hut}}{{Heggie} \&
  {Hut}}{2003}]{heggie03}
{Heggie} D.,  {Hut} P.,  2003, {The Gravitational Million-Body Problem: A
  Multidisciplinary Approach to Star Cluster Dynamics}

\bibitem[\protect\citeauthoryear{{Husser} et~al.,}{{Husser}
  et~al.}{2016}]{spexxy}
{Husser} T.-O.,  et~al., 2016, \mn@doi [\aap] {10.1051/0004-6361/201526949},
  \href {https://ui.adsabs.harvard.edu/abs/2016A&A...588A.148H} {588, A148}

\bibitem[\protect\citeauthoryear{{Ivanova}, {Belczynski}, {Fregeau}  \&
  {Rasio}}{{Ivanova} et~al.}{2005}]{ivanova05}
{Ivanova} N.,  {Belczynski} K.,  {Fregeau} J.~M.,   {Rasio} F.~A.,  2005,
  \mn@doi [\mnras] {10.1111/j.1365-2966.2005.08804.x}, \href
  {https://ui.adsabs.harvard.edu/abs/2005MNRAS.358..572I} {358, 572}

\bibitem[\protect\citeauthoryear{{Kamann}, {Wisotzki}  \& {Roth}}{{Kamann}
  et~al.}{2013}]{pampelmuse}
{Kamann} S.,  {Wisotzki} L.,   {Roth} M.~M.,  2013, \mn@doi [\aap]
  {10.1051/0004-6361/201220476}, \href
  {https://ui.adsabs.harvard.edu/abs/2013A&A...549A..71K} {549, A71}

\bibitem[\protect\citeauthoryear{{Kamann} et~al.,}{{Kamann}
  et~al.}{2018}]{kamann18}
{Kamann} S.,  et~al., 2018, \mn@doi [\mnras] {10.1093/mnras/sty1958}, \href
  {https://ui.adsabs.harvard.edu/abs/2018MNRAS.480.1689K} {480, 1689}

\bibitem[\protect\citeauthoryear{{Kamann} et~al.,}{{Kamann}
  et~al.}{2020}]{Kamann2020}
{Kamann} S.,  et~al., 2020, \mn@doi [\mnras] {10.1093/mnras/stz3583}, \href
  {https://ui.adsabs.harvard.edu/abs/2020MNRAS.492.2177K} {492, 2177}

\bibitem[\protect\citeauthoryear{{Marino}, {Przybilla}, {Milone}, {Da Costa},
  {D'Antona}, {Dotter}  \& {Dupree}}{{Marino} et~al.}{2018}]{marino18}
{Marino} A.~F.,  {Przybilla} N.,  {Milone} A.~P.,  {Da Costa} G.,  {D'Antona}
  F.,  {Dotter} A.,   {Dupree} A.,  2018, \mn@doi [\aj]
  {10.3847/1538-3881/aad3cd}, \href
  {https://ui.adsabs.harvard.edu/abs/2018AJ....156..116M} {156, 116}

\bibitem[\protect\citeauthoryear{{Sabbi} et~al.,}{{Sabbi}
  et~al.}{2020}]{sabbi20}
{Sabbi} E.,  et~al., 2020, \mn@doi [\apj] {10.3847/1538-4357/ab7372}, \href
  {https://ui.adsabs.harvard.edu/abs/2020ApJ...891..182S} {891, 182}

\bibitem[\protect\citeauthoryear{{Sana} et~al.,}{{Sana}
  et~al.}{2013}]{sana2013}
{Sana} H.,  et~al., 2013, \mn@doi [\aap] {10.1051/0004-6361/201219621}, \href
  {https://ui.adsabs.harvard.edu/abs/2013A&A...550A.107S} {550, A107}

\bibitem[\protect\citeauthoryear{{Sun}, {Li}, {Deng}  \& {de Grijs}}{{Sun}
  et~al.}{2019}]{sun19}
{Sun} W.,  {Li} C.,  {Deng} L.,   {de Grijs} R.,  2019, \mn@doi [\apj]
  {10.3847/1538-4357/ab3cd0}, \href
  {https://ui.adsabs.harvard.edu/abs/2019ApJ...883..182S} {883, 182}

\bibitem[\protect\citeauthoryear{{Weilbacher} et~al.,}{{Weilbacher}
  et~al.}{2020}]{pipeline}
{Weilbacher} P.~M.,  et~al., 2020, \mn@doi [\aap]
  {10.1051/0004-6361/202037855}, \href
  {https://ui.adsabs.harvard.edu/abs/2020A&A...641A..28W} {641, A28}

\makeatother
\end{thebibliography}








\bsp	
\label{lastpage}
\end{document}